# Trapping the carrier in the spin-locked $MoS_2$ atomic valley by absorption of chiral L-cysteine

**Susmita Bhattacharya**[*]

*Department of Physics, Indian Institute of Science Bangalore. 560012. India.*

*\*Present address: Radiation Lab, University of Norte dame, Indiana, 46656, USA*

Corresponding Author

[*]**Email: bhattacharyasusmita7@gmail.com**

**Abstract:**

This work, demonstrate enhanced valley contrasting spin-momentum locked chiral states at van der Waal interface of chiral L-cysteine and mono-atomically thin $MoS_2$ placed at $Si/SiO_2$ substrate at ambient condition. Helicity dependent photoluminescence and resonance Raman measurement highlights spin-locked transitions for chiral L-cysteine modified SL $MoS_2$ at ambient condition. Selective adsorption of chiral L-cysteine dimer /cystine stabilise the in-plane effective magnetic field due to $Si/SiO_2$ substrate and blocks the intervalley spin relaxation. The observed polarisation efficiency will be useful for improving the functionality of valley-based light emitting diode (LEDs) and encoding information in logical devices.
.

**Keywords**:  **$MoS_2$, monolayer, chiral molecule, substrate effect, helicity-resolve Raman**,

In a chiral/non-centrosymmetric structure, valley is a quantum number which defines the electronic system using energetically degenerate energy bands with non-equivalent local minima (conduction band) or maxima (valence band). In 2D hexagonal crystals, such as single layer (SL) $MoS_2$, where inversion symmetry is broken, two degenerate valleys can be distinguished by a pseudovector quantity such as Berry curvature and magnetic moment taking opposite values at time reversal pair of valleys[1,2]. The valley contrasted Berry curvature and magnetic moment can couple to external electric and magnetic fields. Because of inversion symmetry breaking, optical interband transitions at time reversal pair of valleys can have valley dependent selection rules. The concept of valleytronics deals with the use of valley index of the carriers (i.e. spin and pseudospin) that process information[3]. The strong spin-orbit coupling in transition metal element give rise to an effective interaction between the valley pseudospin and spin, helping a possible interplay between these two degrees of freedom and allowing spin manipulation via the valley phenomena [2]. It was found that pseudospin polarisation is encoded in the corresponding wavefunction and dipole vector of the optical matrix element [4]. The relaxation of exciton valley pseudospin can also arise from the exchange coupling between its electron and hole constituents. Therefore, this valley dependent phenomena leads to possible manipulation of valley pseudospin by electric, magnetic and optical means. Thus exploring these internal quantum degrees of freedom of carriers will lead us to ascertain their potential usage for new generation electronics for storing and processing more information.

The valley depolarisation phenomena in SL-$MoS_2$ can be viewed by photoluminescence/radiative transitions from excitons and trions as well as non-radiative phonon or defect assisted emission. Valley polarisation, the selective population of one valley can be achieved by tuning the incident photon angular momentum. This polarisation can be preserved for longer than 1 ns. In recent literature, full control over the bright− dark splitting has been described as a pathway to manipulate and control the exciton valley pseudospin dynamics, thus the associated valley polarisation. Designing Coulomb forces between electron and hole and their exchange interaction also has been described as a useful for the same. In SL-$MoS_2$, the lifetimes of the excitons and trions become larger when immersed in high-κ dielectric environments[5]. Coulomb potential becomes more confined within the middle layer (i.e. SL $MoS_2$) if the static dielectric constants of the top and bottom layers are higher. This may help to reduce the scattering between the excitons (trions) and the charged impurities at the $MoS_2$/dielectrics interface, and thus leading to prolong lifetime of the excitons (trions). Modification in interfacial-interaction [6,7,8] between 1L-TMD and preferred molecules is a familiar way to modulate carrier concentration as it does not affect the crystal structure. The spectral weight between excitons and trions of 1L-TMDs can be fine-tuned by controlling carrier densities of 1L-TMDs by means of electrical/chemical doping [9]. Calculation of spin density in presence of n-or p-type of dopant clarifies that p-orbital of the dopant plays important role in modulating electronic and magnetic property of layered MoS$_2$, although the role of d-orbital of Mo is the most significant [10]. Here it is to be noted that different surface treatments of SL-MoS$_2$ has been reported only to improve the photoluminescence but their influence on the valley polarization has not yet been explored till date.

Highly polarised valley singlet and triplet interlayer excitons are found in van der Waal heterostructures[11] following electron-hole separation and discretised singlet and triplet transition. Spectroscopically, a spin singlet exciton has in-plane transition dipole [12]. It couples to σ+ (σ−) polarised photon (left and right circularly polarised respectively) propagating in the out-of-plane (z) direction. On the contrary, the weak radiative recombination of a spin-triplet exciton which is generally dark, can emit a z-polarised photon propagating in the in-plane direction [13]. Brightening of dark excitons by imposing in plane magnetic field [14] was found as a possible strategy to enhance the population of excitons and enhanced valley polarisation. The population decay of exciton following different radiative and non-radiative process that can be utilised to understand the extent of valley polarisation and the preferred decay channels. Unexpectedly, at low carrier density, substrate induced localised in-gap states degrades valley polarisation in 2D layered material[15]. At room temperature the decay in valley-based photoluminescence helicity is mostly governed by the electronic spin relaxation, phonon-assisted [16,17] and defect assisted scattering phenomena [18]. At ambient condition the emission spectra is mostly governed by negative excitonic transition. The spin relaxation of negatively charged exciton becomes equivalent to intervally relaxation [1] and the possible spin relaxation mechanisms for a charge carrier in a semiconductor is considered to controls the relaxation of valley spin polarisation in SL MoS$_2$. Therefore, the manipulation of valley spin by coupling it with electron spin of chiral molecule of preferred symmetry will be an ingenious way to achieve efficiently enhanced valley polarisation in SL MoS$_2$ at ambient temperature.

In this study, the modified spin relaxation mechanism at the chiral L-cysteine -SL MoS$_2$ van der interface at ambient condition. The extent of valley polarisation in pristine SL MoS$_2$, SL MoS$_2$ when attached with chiral/ achiral molecule (Ch-SL/aCh-SL) is clarified by considering increment in photoluminescence helicity contrast (**ρ**). The possible depolarisation pathways explored by helicity-resolved Raman spectra will further confirm the manipulation of spin/pseudospin in SL-MoS$_2$ valley exotic states in presence of chiral L-cysteine molecule at ambient condition.

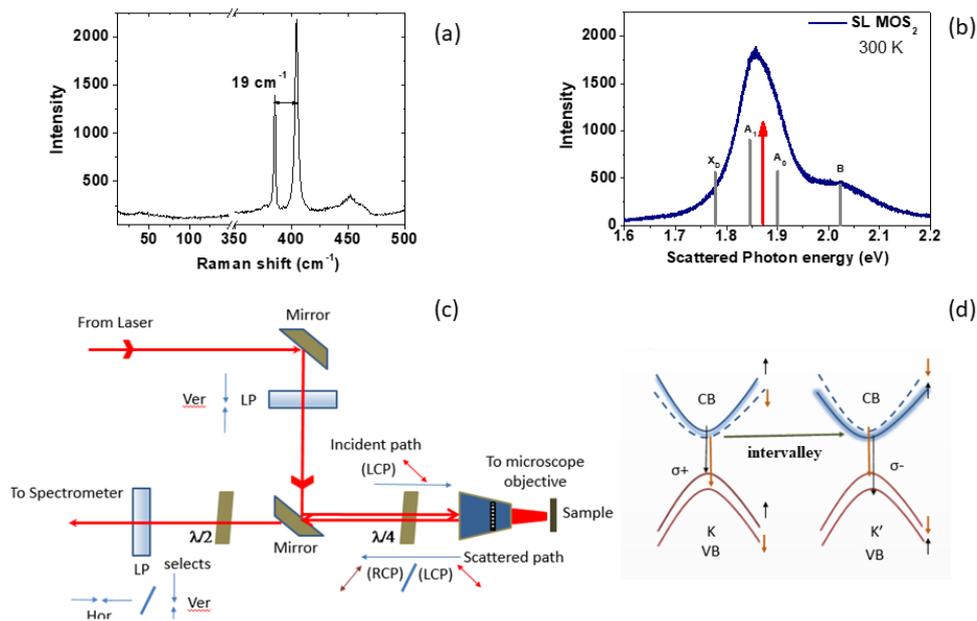

**Figure 1. Primary feature of monolayer MoS$_2$, a) Raman spectra of monolayer MoS$_2$, b) Photoluminescence spectra of pristine MoS$_2$ when excited by 2.32eV at 300K (Nevy blue color line), gray color lines describe peak position of A1: trion, A0: exciton, B: exciton). Red arrow line is marked as the resonance excitation used to probe the sample property. c) Optical setup for helicity resolved spectroscopy. The 2$^{nd}$ column in table S1 describes the detection configuration. (d) Schematic representation representation of valley-based transitions.**

**Experimental Details**

Monolayer MoS$_2$ were mechanically exfoliated from bulk MoS$_2$ (SPI supplier) and were placed on SiO$_2$/Si surface [details of exfoliation procedure mentioned in Section S1 of the supplementary materials, including the AFM image and height profile in Figure S1(a,b)]. The location and thickness of the flake were identified by Raman spectroscopy and PL characteristics feature (Figure 1, a and b respectively). The sample was then annealed at 250°C for 1hr. For attachment of achiral cysteamine and chiral L-cysteine on the SL MoS$_2$ surface, the MoS$_2$ sample on SiO$_2$/Si substrate was soaked in 1/40 v/v Cysteamine/methanol and L-Cysteine/isopropyl(IPA)-water solution for 72hrs and 24 hrs respectively. The samples were then rinsed thoroughly with ethanol and IPA to remove unbound molecules. As the deposition of thick organic layer degrade device performance, thus the above-mentioned protocol is followed to get optimised concentration of equal length chiral L-cysteine and achiral cysteamine molecules having O /S terminal are localised at the S-defect sites.[9, 19, 20] Figure S2 in supplementary section show that achiral Cysteamine and chiral L-cysteine molecules were tightly bound on the surface of MoS$_2$ (aCh-SL and Ch-SL) at their optimal concentration after the repeated washing.

The helicity-resolved Photoluminescence (PL) and Raman measurement were performed with a micro-Raman Spectrometer (M/S Horiba) equipped with a peltier cooled CCD. The optical setup has been described in Figure 1(c). The excitation laser was first guided through a vertical linear polarizer followed by a quarter-wave plate to achieve σ+(left) circular polarization. The circular polarization of the excitation light was confirmed at the sample position. The backscattered signal passes through the same quarter-wave plate is collected and analysed with a half-wave plate and a linear polarizer. Rotation of the half-wave plate at different angles enables us to select the helicity (ρ) of the scattered light. σ- polarisation of incident light gives similar selectivity. But for clarity results with σ+ polarisation of incident light will be discussed. Two different laser sources are used for the measurements: Nd:YAG solid state laser (532 nm, 2.32 eV), and red laser (660 nm, 1.88 eV) with incident fluence of $1.72\times10^{-2}$ and $3.4\times10^{-2}$ W/m$^2$. The spectral resolutions are 0.45 cm$^{-1}$ and 0.35 cm$^1$ for helicity resolved Raman experiments using 1800 lines/mm grating. In the text, the degree of PL polarisation has been quantified by the helicity contrast and will be discussed later. For low temperature measurements, a Linkam THMS-350 heating and freezing stage was used to obtain temperature variation up to 78K. All the PL, and Raman spectra are collected at ambient condition, 300K and 78K in two different detection configurations (parallel and cross). The Raman shift were calibrated using the Si Raman peak at ~520 cm$^{-1}$ for both detection configuration.

To confirm the quality of our pristine sample, Raman and PL measurements using excitation of 2.32 eV were employed as shown in the Figure 1, panel a, b. Monolayer MoS$_2$ is a direct band gap semiconductor with energy gap located at K and K′ points of the Brillouin zone where the highest and lowest parts of corresponding valence and conduction bands have significant contribution of Mo d-orbitals and S p-orbitals. The valence band is split by ~150 meV due to the spin-orbit interaction. Raman active $A_{1g}$ (405 cm$^{-1}$) and $E_{2g}^1$ (386 cm$^{-1}$) modes in backscattering geometry are widely used to measure the layer thicknesses and crystal qualities. The difference in Raman shift for $E_{2g}$ and $A_{1g\ modes}$ ~ 19 cm$^{-1}$ and PL characteristics spectra for monolayer MoS$_2$ as shown in Figure 1 (a) and (b) confirm characteristics of pristine monolayer MoS$_2$ recorded in the ambient condition. The peak positions marked by grey color lines in PL spectra of Figure 1(b) signifies intense peak position related to trion (A1) due to unintentional n-doping of MoS$_2$ sample, exciton A0 and exciton B at 300 K. The peak position of exciton bound to defect, X$_D$ is also marked by grey vertical lines, the peak position corresponding to exciton is ~1.9eV and for trion is 1.84 eV for trion [21]. Upon changing the temperature to 78 K, the maximum of the luminescence peak shift to higher energy. With decrease in temperature the excitonic peak sifts to 1.96 eV and trion peak shifts to 1.87 eV. Our excitation energy, 1.88eV (marked by red arrow) is in resonance with the neutral exciton A0 at 300K and in resonance with the trion at 78 K [data not shown].

Figure 1 (c) describes a schematic diagram of our experimental setup. The atomic layer of MoS$_2$ on Si/SiO$_2$ substrate is excited with left circularly polarised light (σ+) of excitation energy 1.88eV (in resonance with A0) and detected separately for σ+ (left) and σ- (right) emission configuration for all three types of samples. Table SI in supplementary section describes the detection configuration. In figure 1(d), schematically describe the population decay in SL-MoS$_2$ of neutral and charged excitons to photons, intravalley and the intervalley relaxation at steady state. The valence band and conduction bands in SL MoS$_2$ are split into four bands although the conduction band (CB) splitting is very small. For singlet exciton, the hole in the lower state has the same spin as the electron in CB while for triplet exciton the it is opposite. Thus, in SL MoS$_2$ the excitonic transition is normally singlet in nature although triplet transitions are induced by conduction band splitting

The modifications in the atomic layer of MoS$_2$ by adsorption of chiral and achiral molecules are clarified by off-resonance (2.32eV) PL (a,b) and Raman spectrum (c,d) Figure S1 in supplementary section. A shift and decrease in luminescence signify n-doping for aCh-SL system while an enhancement in luminescence for Ch-SL suggests p-doping. In Figure S1 (c, d), the observed shift in A$_{1g}$ and E$^1_{2g}$ for attachment of both the molecules is found to be negligible (~0.5 cm$^{-1}$) referring the doping level to be ~10$^{12}$/cm$^2$. The unpolarized emission with off-resonance 2.32 eV excitation for all the three samples i.e.(a) SL-MoS$_2$, (b)Ch-SL, (c) aCh-SL, are shown in Figure S2, has been interpreted as an indication of loss of valley polarization due to simultaneous population of both the valleys[1] for all the three samples.

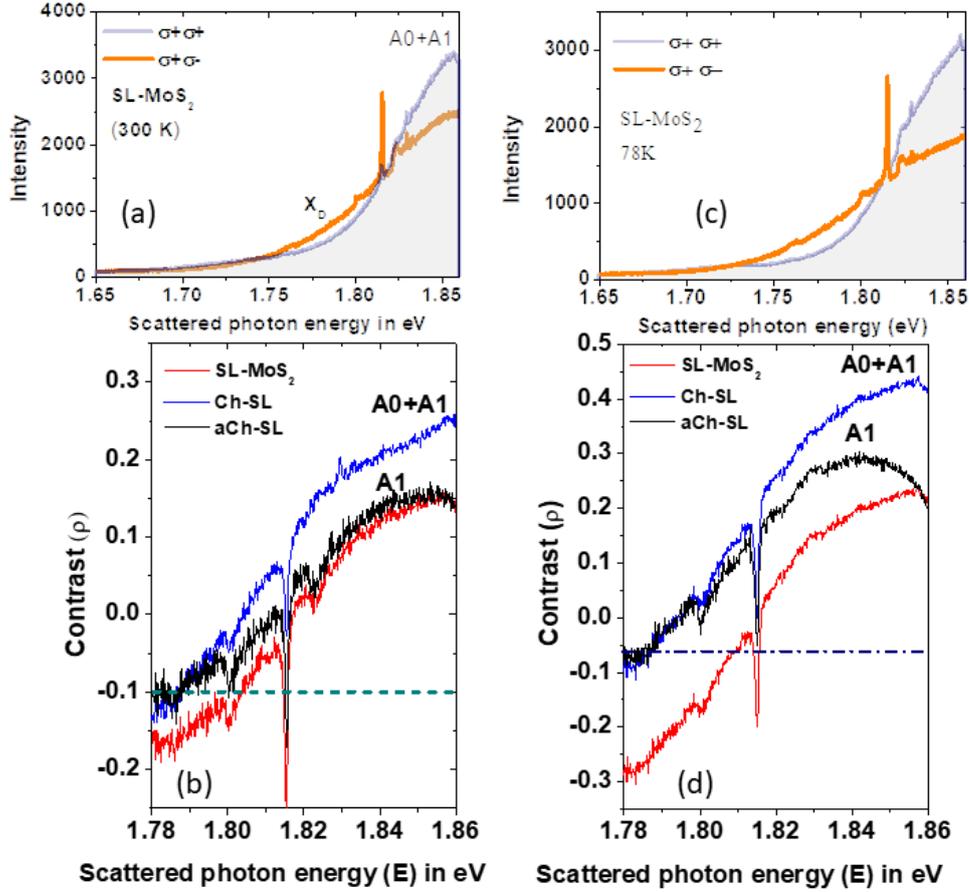

**Figure 2:** (a) Photoluminescence spectra of pristine SL-MoS$_2$ at 300 K with incident left circularly polarised light (σ+) excitation of 1.88 eV. Violet line corresponds to same detection configuration (σ+) and orange line cross detection (σ-) configuration. The area under the curve for same detection configuration describes the enhanced population of carrier upon excitation with circularly polarised light, while the cross detection defines the relative loss in population. (b)The helicity contrast spectrum for SL MoS$_2$ (red), Ch-SL (blue) and aCh-SL (black line). (c) Photoluminescence spectra of pristine SL-MoS$_2$ at 78 K with incident left circularly polarised light (σ+) excitation of 1.88 eV. Blue line corresponds to same detection configuration (σ+) and orange line cross detection (σ-) configuration. (d) Corresponding helicity contrast spectrum for SL MoS$_2$ (red), Ch-SL (blue) and aCh-SL (black line) at 78K.

**Helicity resolved photoluminescence tuning incident photon angular momentum polarised light**

The connection between valley degree of freedom and photon helicity in SL MoS$_2$ is understood on angular momentum conservation [22] mechanism. Thus, the valley-selective excitation near the K (K′) point is also the spin-selective excitation when the excitation laser energy matches to the energy gap. This fact makes it possible to detect the valley polarized electron by the direction of spin. The spin relaxation of a singlet negatively charged exciton reflects the relaxation of a hole. In all three samples MoS$_2$ under study, SL, aCh-SL, Ch-SL, hole spin relaxation is a measure of depolarisation. Here, it is considered that for holes, valley and spin indices are locked at the valence band Bloch states. The σ+ excitation creates excitons with electron spin up and hole spin down at the K point. In literature, it is claimed that the valley depolarization originates from the electron/hole spin relaxation due to the D'yakonov-Perel' (DP) and Elliott-Yafet (EY) [23] and BAP mechanisms. PL depolarization also includes intervalley scattering including the electron-phonon and/or short-range impurity scatterings, the spin relaxation of the electron and hole can even cause the bright exciton transition between the *K* and *K′* valleys. Intravalley scattering is forbidden due to the spin degeneracy near the valence band edge, splited by ~150 meV. Again, intervalley scattering from the K to K$^{'}$ point involving simultaneous spin flip requires coupling with both atomic scale scatterer and with magnetic defects [12]. Figure 2 (a) and (c) describes the optical control of valley spin polarisation in SL MoS$_2$, nearly in resonance with A0 (exciton) and A1(trion) at ambient and low temperature. A steady state, helicity-resolved photoluminescence spectra with σ+ excitation for pristine SL in ambient condition (300K) has been depicted in Figure 2(a). The dark blue and orange colour line correspond to parallel and cross detection configuration of the spectrum. The area under the curve describe the population of excitons and trions in the corresponding configuration. The observed steady state PL spectra is mainly governed by the valley polarisation during the valley carrier initialisation process. In same detection configuration (σ+σ+) valley selective carrier relaxation can be observed, while in cross configuration (σ+σ-) the valley selectivity is lost. The extent of valley polarisation with variation in scattered photon energy, for these three different systems helicity contrast (**ρ** ), as

$$\rho=(I_+ - I_-)/(I_+ - I_-) = (n_k - n_k')/(n_k + n_k')\ldots\ldots..1)$$

where, $I_+$ and $I_-$ are the intensities of the PL signals corresponding to the same (σ+σ+) or cross (σ+σ-) circular polarisation with respect to incident polarization[1] while $n_k$ and $n_k'$ are the population of excitons in K and K′ i.e hole spin up and hole spin down valleys. Here it is considered that for holes, valley and spin indices are locked at the valence band Bloch states. Figure 2 (b) describes the spectral variation in helicity contrast ($\rho$) at 300 K for all the three samples SL (red line), Ch-SL (blue line) and aCh-SL (black line). For SL, spectrum i.e. ($\rho$) vs scattered photon energy (E) in eV plot, consists of equal dominance of luminescence of A0 and A1 peak at 1.86 eV in the σ+ polarization detection configuration but the extent of polarisation is found to be more for Ch-SL sample ($\rho$ ~0.25). A very weak emission from defect trapped exciton, $X_D$ at 1.77 eV [24] is observed for σ- polarization detection configuration SL sample. For aCh-SL the helicity contrast peak is lower and red shifted to 1.84eV due to more contribution from trion A1. Here it is needed to be mentioned that intimate proximity to surface defects, dangling bonds, dielectric disorder, and surface roughness of the Si/SiO$_2$ substrate[15], and effective temperature degrades the valley polarisation of the MoS$_2$ sample. Figure 2(c) describes the helicity resolved PL spectra for SL at 78 K and subsequent helicity contrast spectra for all three sample are shown in Figure 2(d). For comparison between helicity contrast ($\rho$) for valley exciton A0, trion A1 and defect exciton $X_D$ observed specially at two temperatures (300K, 78 K), a parallel blue dashed line is defined in Figure 2.b) and d) as a guide to the eye to estimate the extent of polarisation. We have observed at 78 K, $\rho$ value for A exciton is higher for Ch-SL sample at 1.86eV compared to the pristine. Spectrum for aCh-SL portrays an enhanced trion helicity ($\rho$ ~0.3 compared to pristine A exciton $\rho$ ~0.25) at 1.84eV. As with decrease in temperature the photoluminescence peak blue shifts, maximum extent $\rho$ was far higher energy of our detection limit (1.88eV). With decreasing temperature, as described in figure 2d), $\rho$ is maximised for Ch-SL at scattered photon energy of 1.86 eV. However, it is observed that defect passivated aCh-SL leads to small enhanced helicity contrast ($\rho$) around 1.84 ev (trion A1) at 78K as shown by black line in Figure 2(d).

According to BAP mechanism, spin relaxation through the e-h exchange interaction are prominent for 2D material when all other mechanisms are freezed. The presence of such an interaction allows a valence hole spin in lightly n-doped MoS$_2$ monolayer to relax through a simultaneous valley- and spin-flip scattering with the conduction band electrons i.e intervalley scattering. It does not include any intravalley phenomena or spin flip phenomena for the same valley. It includes an exchange splitting on the order of 1 meV, an unintentional doping level of $5\times10^{12}$ cm$^{-2}$, which is true for all three samples under observation. Thus, BAP mechanism appears to be an effective relaxation process in monolayer MoS$_2$ even at 78K for all three samples for favouring the excitonic singlet transition. The other mechanisms, such as DP and phonon mediated scattering contribution becomes minimum at this thermal energy. This claim will be further clarified by helicity resolved Raman scattering. Other than the above-mentioned mechanism the substrate induced effect also plays a critical role defining valley depolarisation. In SL MoS$_2$, the exact 2D motion of the charge carriers and the presence of mirror symmetry plane in the $D_{3h}$ point group imply that the effective magnetic field, felt by the spin of a charge carrier has no in-plane component. Mirror symmetry, in SL MoS$_2$, can be broken in substrate-supported samples or in field-effect transistor structures in which electric fields may be present. It could lead to spins of charge carriers with different crystal momenta to precess at different rates between scattering events with possible connection between acoustic phonons LA associated with K, K′ points.

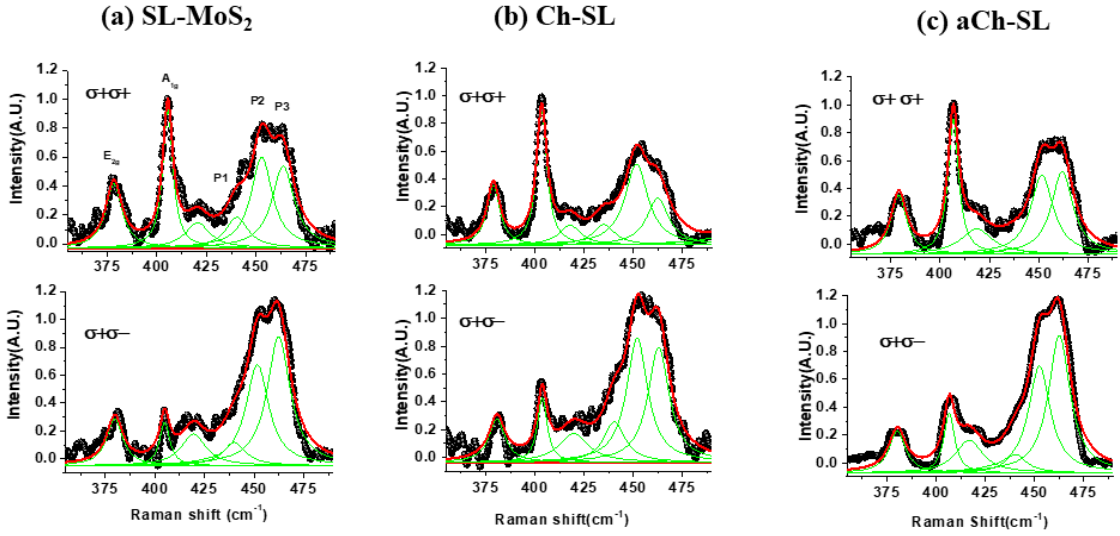

**Figure 3: Deconvoluted Raman spectra of pristine (a)SL-MoS$_2$, (b) Ch-SL, (c)aCh-SL with 1.88eV laser at 300 K in two detection configurations after subtraction of excitonic background.**

Resonance Raman scattering of different phonon modes with polarized incident radiation reveal information of electronic states with specific symmetries, present at the band edge which cannot be extracted with non-resonance excitation. When circularly polarised light is incident, resonant Raman intensity of the first order bands of pristine MoS$_2$ depends on electron-photon and electron-phonon matrix elements related to the initial and final state[16,17]. Contrary to a first order Raman band, a second order Raman band can be expressed as a convolution of multiple two phonon process across the Brillouin zone and very often the scattering process are found to be doubly degenerate [25].

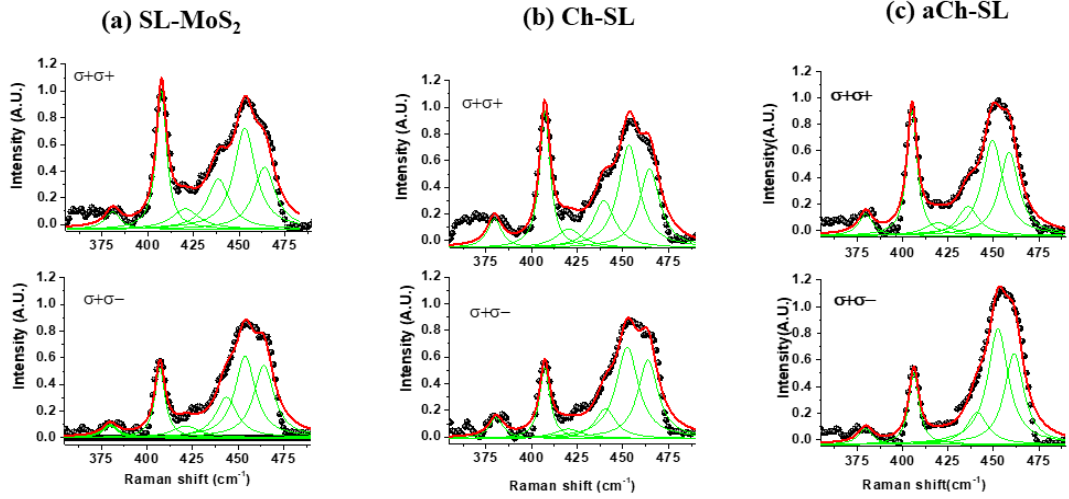

**Figure 4: Deconvoluted spectra of pristine (a) SL-MoS$_2$, (b) Ch-SL, (c) aCh-SL with 1.88eV laser at ambient 78K in two detection configurations after subtraction of excitonic background.**

Here, Figure 3 (a, b, c) describes the resonance Raman spectra of SL, Ch-SL and aCh-SL at ambient temperature (300K) while Figure 4 (a, b, c) describes the spectra collected at 78K when excited with 1.88eV excitation. For incident light σ+ polarisation and detection in σ+/σ- (same and cross) configuration, the obtained feature are compared with the selection rules in Table S1 as per Ref 16. The A$_{1g}$ mode is found to be present for all three samples, with its maximum intensity for σ+ polarisation configuration at resonance i.e. 1.88 eV and off-resonance 2.32 eV excitation energy (data not shown), and it agrees with Raman selection rules for all three samples at room temperature. For E$_{2g}$ mode, the behaviour is different for excitation with 1.88eV compared to the off-resonance condition[26]. The crystal symmetry forbids the E$_{2g}$ mode to be present at parallel configuration, and permits to be present in cross configuration with maximum intensity. Contrary to this understanding, E$_{2g}$ mode is found to be present for all three types of samples for same incident and scattered polarisation at ambient condition (300K) as shown in Figure 3 (a, b, c). Moreover, at 78K, the intensity of the same mode got lowered even in cross polarisation configuration as presented in Figure 4 (a, b, c). The phenomena can be interpreted as lesser coupling to in plane mode E$_{2g}$ mode with A exciton [27, 28]. A exciton transition connects lower part of conduction band composed of Mo d$_z^2$, for the electron and upper part of valence band of d$_{xy}$ character for the hole. Thus, stronger electron-phonon coupling is expected for out of plane A$_{1g}$ vibration and as a result A$_{1g}$ feature is enhanced in comparison to in plane $E_{2g}$ mode when the excitation laser is in resonance with A exciton. Another interesting feature that attracts attention is that broad acoustic mode around 450 cm$^{-1}$ shows a prominent change in lineshape when detected in cross polarisation at 300K. At low temperature, 78K, the band lineshape remains unaltered for both detection configuration for all three samples.

The resonant excitation reduces the number of possible spin relaxation mechanisms such as coupling to the excited states. It is believed that a threshold of twice the LA phonon energy, specific to the material, above which phonon-assisted intervalley scattering are found to cause depolarization[17]. There are two possible mechanisms, responsible for the electron or hole spin-flip during this phonon mediated intervalley scattering event. One is that the spin-flip is mediated by short range scattering from impurities. The presence of a background carrier population could enhance the probability of such a process [22]. The other mechanism is that intervalley scattering proceeds through the nearly spin-degenerate Γ valley of the Brillioun zone (BZ) [29]. Following this way, Large-momentum low-energy exciton states can provide relaxation channels for bright excitons, and reduce photoluminescence quantum efficiency. For SL MoS$_2$, the 1$^{st}$ order bands are related to estimation of polarisation [16,22] and the second order double resonance spectra of MoS$_2$ originates due to intervalley scattering by acoustic phonon, mechanism responsible for destruction of valley polarisation [17].

To estimate the extent of polarisation and the proper decay channel responsible for valley depolarisation, the spectral range 355 cm$^{-1}$ to 490 cm$^{-1}$ has been deconvoluted with a sum of Lorenztian functions after subtracting excitonic background. Figure 3 (a,b,c) describes the deconvoluted spectra for SL-MoS$_2$, Ch-SL and aCh-SL where the contribution of $E_{2g}$ (380cm$^{-1}$), $A_{1g}$ (406 cm$^{-1}$), a side LAside/LAs band (419 cm$^{-1}$), and other three second order bands are considered. In resonance condition, higher order combination bands and zone edge acoustic phonons near M and K points of the Brillouin zone, such as 2LA (K/M) ~ 450 cm$^{-1}$, are seen along with the luminescence background. Three Lorentzian functions (notation P1~436 cm$^{-1}$, P2~452 cm$^{-1}$, P3~462 cm$^{-1}$, width~14 cm$^{-1}$) are used to describe the broad spectral feature around 450 cm$^{-1}$. The non-dispersive 2$^{nd}$ order P2 mode is related to K and M point in BZ while dispersive P3 is related to scattering process from K to K′. Si peak at 520 cm$^{-1}$ has been considered as internal standard to calibrate the frequency shift. During deconvolution procedure, the intensity, width, peak position are considered as free fitting parameter as mentioned in Ref 17.

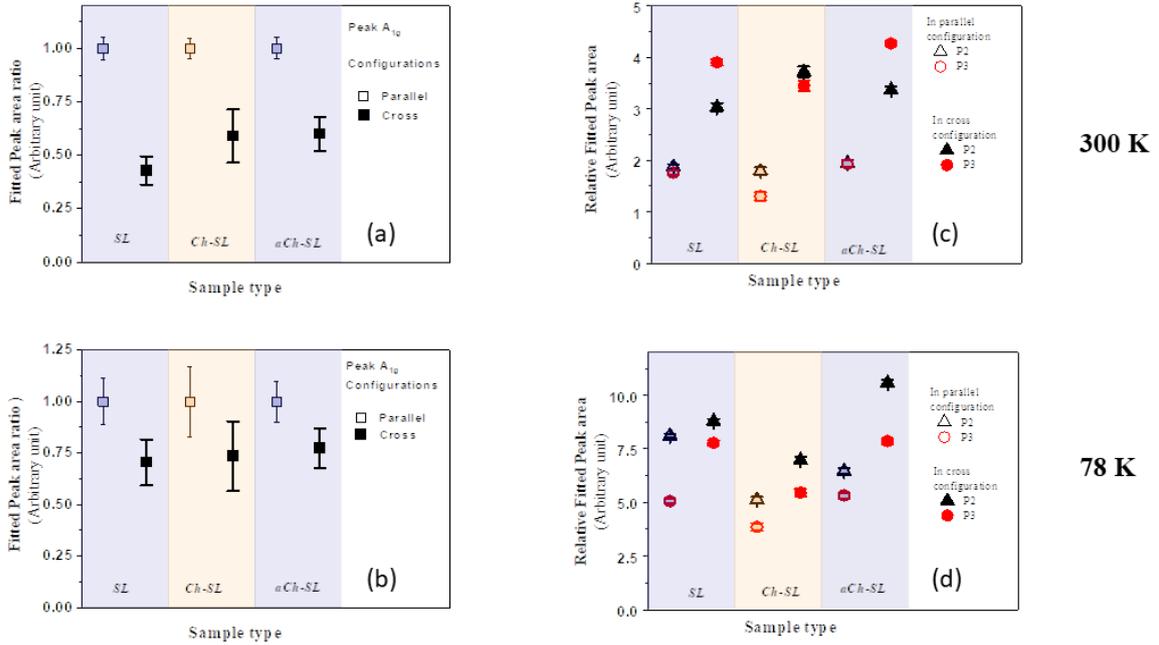

**Figure 5. a) Comparison of evolution of $A_{1g}$ optical mode at (a) 300K and (b) 78K in two detection configurations (respect to $E_{2g}$ mode, considering intensity of $A_{1g}$ mode maximised in parallel component of emission) for three different samples. Evolution of P2 and P3 mode for three different samples at (c) 300 K and (d) 78K with respect to $E_{2g}$ mode.**

At 78K, the same fitting procedure is followed for same spectral range of three different samples, considering the hardening of Raman mode. As $E_{2g}$ peak height remains unaltered upon two detection configuration at 1.88eV, it has been used for intensity calibration.

In a recent work by Drapcho *et. al* [26], appearance of $E_{2g}$ mode in same polarisation was related to a defect-assisted process that involves phonons in the transverse optical E branch slightly away from Γ the point. This process can be enhanced by the selective coupling of valley pseudospin to photon helicity. The similar results i.e. presence of $E_{2g}$ mode in parallel polarisation is also observed. Thus, the valley pseudospin, in addition to the crystal symmetry, plays a significant role in understanding the resonance Raman scattering spectra for excitations close to valley exciton resonances (A exciton here). At 78 K, intensity of $E_{2g}$ mode compared to $A_{1g}$ is lowered for all of the three samples as shown in Figure 4 (a,b,c) due to change in resonance condition which is otherwise prominent at room temperature for all the three samples. Next, the extent of valley polarisation following the coupling with out of plane $A_{1g}$ mode with exciton/trion and depolarisation pathways, non-dispersive 2$^{nd}$ order P2 mode connecting K and M point in BZ; dispersive P3, related to scattering process from K to K′ at ambient (300K) and low temperature (78K) will be discussed.

Figure 5(a) and (c) describes the change in intensity of A$_{1g}$ mode in two detection configurations (open symbol for same, filled symbol for crossed configuration) at 300K and 78K respectively. The same detection configuration is considered to define intravalley relaxation mechanism while the cross-detection configuration defines intervalley spin relaxation mechanism. It gives an idea about the light coupling efficiency for the extent of valley polarisation that can be reached for the mode upon 1.88eV excitation. At 300K, the intensity of out of plane A$_{1g}$ mode (mostly related to Γ points in BZ) is lowered at cross configuration for SL-MoS$_2$ compared to Ch-SL and aCh-SL samples. Adsorption of achiral cysteamine/chiral cysteine on MoS$_2$ modifies the fast relaxation via optical phonon near K point at ambient condition, and as a result, the A′ or A$_{1g}$ mode (other than Γ points in BZ) mode contributes in cross polarisation configuration. At 78 K (Figure 5(c)) as the laser excitation energy is in resonance with trion peak and the excess electron-phonon contribution becomes prominent for all three samples and A$_{1g}$ mode intensity remains unaltered at cross polarisation.

Next, we look into, the relative intensity ratio of P2 and P3 mode (with respect to E$_{2g}$ mode for each of the sample) with in Figure 5 (b) and (d) for 300 K and 78 K. At 300K, in parallel configuration, higher value for P2 mode compared to P3 mode is observed as shown in Figure 5(b). In cross configuration i.e. while probing the intervalley scattering, P3 mode is found to be dominated over P2 mode for SL, aCh-SL. For Ch-SL, in both the configuration P2 and P3 mode are found to have similar intensity contribution in cross configuration and cannot overpower contribution from P2 mode. Thus, the effective depolarisation from K to K′ or via other points in the BZ are decreased in case of Ch-SL sample at 300K.

When the variation of non-dispersive P2 and dispersive P3 mode at 78K is considered Figure 5(d), for all three samples P2 mode is dominant over P3 even in cross polarisation configuration. Therefore, it is found that the intensities of K to M transition (intravelly) is much higher compared to K-K′ the intervalley phonon mediated transitions. The dispersive P3 mode disperses rapidly away from K point and the probability of intervalley transition becomes much lower at 78K. Here it should be mentioned that at 78 K, A$_0$ exciton selectivity is lost which is being reflected in the corresponding Raman features as well. The excitation laser line is in resonance with the A1 trion peak at 78K. The DP mechanism (intervalley) fades out only BAP mechanism (intravalley) becomes responsible for spin relaxation. For proper estimation of depolarisation mechanism at 78 K, the samples need to be probed at $A_0$ excitation wavelength

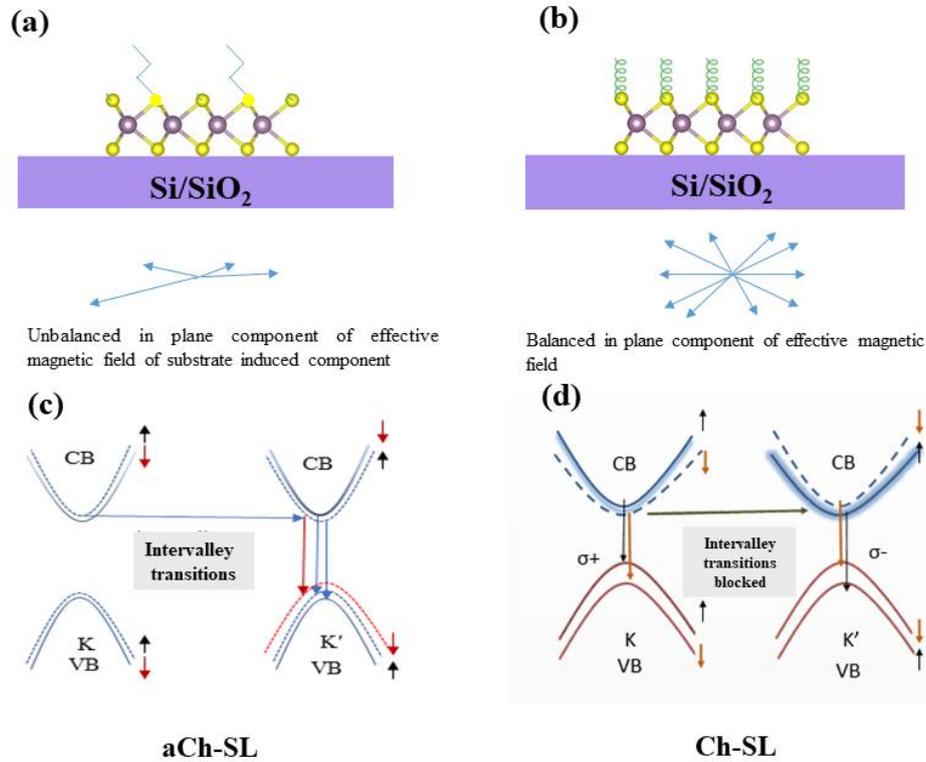

**Figure 6: Schematic representation of (a) unbalanced in-plane component of effective magnetic field in aCh-MoS$_2$ and (b) balanced in-plane component in Ch-MoS$_2$. (c) Intervally transitions (blue bands and blue lines) and presence of in-gap states in aCh-MoS$_2$ (band and transition marked by redline). The intermediate in-gap states with negligible or small in-plane spin-orbit effective fields can mix spins via precession and thereby mediate K→K′ intervalley relaxation. d) Transitions between spin locked states and blockage of intervalley transition in Ch-SL.**

For pristine SL-MoS$_2$ on Si/SiO$_2$ substrate, spin-locked intravalley transitions are lowered in presence of intervalley (possibly K-K′ transitions, transition connecting M and K points in the BZ, presence of in-gap states due to unintentional defect and substrate effect) transitions following relaxation channels via multiple acoustic phonons. In aCh-SL, the s-deficient states are removed but the chemical coordination of the molecule with SL-MoS$_2$, induces a small n-doping effect (~$10^{12}$ cm$^{-2}$) leading to generation of trion weighted spectral feature. While radiatively decaying, trions eject an electron and the recoil electron carries away all of the trion's momentum, allowing all trions to decay radiatively. Thus, including the energy of the recoil electron is essential when determining the PL spectral shape from the trion momentum distribution[19]. In aCh-SL placed on Si/SiO$_2$ substrate, the momenta of charge carriers are randomized by scattering and results in unbalanced the in-plane component of magnetic field and spin lifetimes for holes, and the enhancement in emission is not observed at 300K [15]. It induces more channels for spin relaxation (as observed from enhanced 2LA mode) and reduces the valley polarisation. The enhanced valley contrast (ρ) in presence of chiral L-cysteine molecules is mostly governed by the formation of L-cysteine dimer at pH 10. The formation of dimer is driven by optimisation of each cysteine molecule on MoS$_2$ similar in case of 110 Au surface [30]. It introduces selective adsorption of oxygen (O) at the s-defect site as well as chemically absorbed top of S atom present at the MoS$_2$ basal plane [31]. It can be inferred that adsorption of O at the basal plane of MoS$_2$ balances the substrate induced magnetic field, reduce depolarisation due to in gap states. This adsorption modifies the behaviour of the confined charge carriers in Ch-SL-MoS$_2$ at Si/SiO$_2$ substrate and reduce the unbalanced K dependent components. A schematic illustration is included to describe the absorption procedure in Figure 6 (a, b) and a comparison of valley based transitions in Figure 6 (c, d) at the interface of achiral cysteamine-SL MoS$_2$ and chiral L-cysteine - SL MoS$_2$. The K to K′ intervalley transition and transition connecting M and K points is lowered by preferred locking of spins at the VB states. As a result, an enhanced valley contrast transition in Ch-SL MoS$_2$ sample is observed. Therefore, in this work, a simple and effective method to control the population of spin-locked states of confined carrier in 2D SL-MoS$_2$ by adsorption of chiral L-cysteine is described. A further estimation of exciton/trion g-factor temperature dependence from magneto PL measurement [32,33] will give precise estimation of the spin locked states in the modified atomic valley.

In summary, defect passivation with achiral cysteamine molecule induce n-doping effect, cannot remove in-gap states formed due to Si/SiO$_2$ substrate. The helicity dependent photoluminescence and Raman scattering measurements highlights that balanced in-plane magnetic field due to selective absorption of chiral molecule with spin-orbit effective field component in Ch-SL MoS$_2$. The spin relaxation in the system is modified lowering the intervalley transitions. This gives rise to enhanced valley polarisation Ch-SL MoS$_2$ in ambient condition. Ch-SL MoS$_2$ interface can be utilised as a model system to maximise valley polarisation at ambient condition. We believe that, our experimental understanding can be exclusively applicable for room temperature valley information based LED and photonic logical device.

**Acknowledgement**: S.B. thanks DSKPDF, UGC for financial assistance and Prof. A.K. Sood for constant support and invaluable mentorship. S.B Thanks Srishti Pal and Dr. D.V. S. Muthu, IISc Bangalore for help during the low temperature measurements. S. B. thanks anonymous reviewer for valuable comments.

**Data availability statement:** The data that supports the findings of this study are available within the article [and its supplementary material].

# Supplementary information

Table S1: Proposed symmetry of 1L MoS$_2$ (D$_{3h}$) as obtained from Multiplication table [Ref 16]

|  | Configuration | Symmetry of matrix element Photon | Symmetry of matrix element Phonon | Final |
|---|---|---|---|---|
| IMC mode | z′(σ$_+$σ$_+$)Z | A$_1$′ | E′ | E′ |
|  | z′(σ$_+$σ$_-$)Z | E′ | E′ | A$_1$′+A$_2$′+ E′ |
| OC mode | z′(σ$_+$σ$_+$)Z | A$_1$′ | A$_1$′ | A$_1$′ |
|  | z′(σ$_+$σ$_-$)Z | E′ | A$_1$′ | E′ |

Table S2: Peak positions and corresponding assignments of each Raman mode observed

| SL MoS$_2$ Peak position (cm$^{-1}$) | Assignment |
|---|---|
| 178 | A$_1$′(M)-LA(M) |
| 376 | E′(M) |
| 385 | E′(Γ)/E$_{2g}$ |
| 404 | A′(Γ)/A$_{1g}$ |
| 414 | E′(M) |
| 452 | 2LA(M) |
| 460 | A$_{2u}$ (Γ) |
| 638 | A$_1$′(M)+LA(M) |

## Section S1: Exfoliation procedure

Single-layer MoS$_2$ flakes were mechanically exfoliated from a bulk single crystal procured from M/s SPI Supplies and transferred on a thoroughly cleaned ( step 1: removal of metallic impurities, followed by step 2: RCA based method to remove organic and ionic contamination, step 3: backed ~110°C to remove moisture) 300-nm SiO$_2$/Si substrate to get a good contrast of the flake under optical micrograph. The layer thickness was further confirmed by AFM height profile. The topographic images of samples were measured using an atomic force microscope (AFM, Innova, Bruker) as shown in Figure S1.

## Section S2: Adsorption of achiral and chiral molecule on SL MoS$_2$ surface

In order to understand the effect of the surface charge transfer due to addition of achiral and chiral molecules on the property of charge carriers in MoS$_2$, we have performed PL and Raman measurements. The structure of the corresponding molecules are shown in the inset of Figure S2 (a, b). The doping effects of the molecule can induce shifts in the corresponding spectral features Figure S2(a) depicts that the measured PL spectra of $MoS_2$ after attachment of achiral cysteamine molecule is redshifted and noticeably weakened, while Figure S2(b) clarifies attachment of chiral molecule induces small increase in PL intensity although the shift in peak position is very negligible. We attribute the effect observed for attachment of achiral cysteamine molecule to be electron-doping effect. For looking into the electron-phonon interaction in the modified systems, we looked into Raman spectra of these systems. In Figure S2 (c, d), the observed shift in $A_{1g}$ and $E^1_{2g}$ for attachment of both the molecules is found to be negligible (~0.5 cm$^{-1}$) when excited by 2.32eV excitation energy. The modified systems were found to be stable under ambient condition for 4-7 days.

Figure S3 describes the helicity resolved luminescence spectra for all three samples when excited by 2.33 eV (off-resonance).

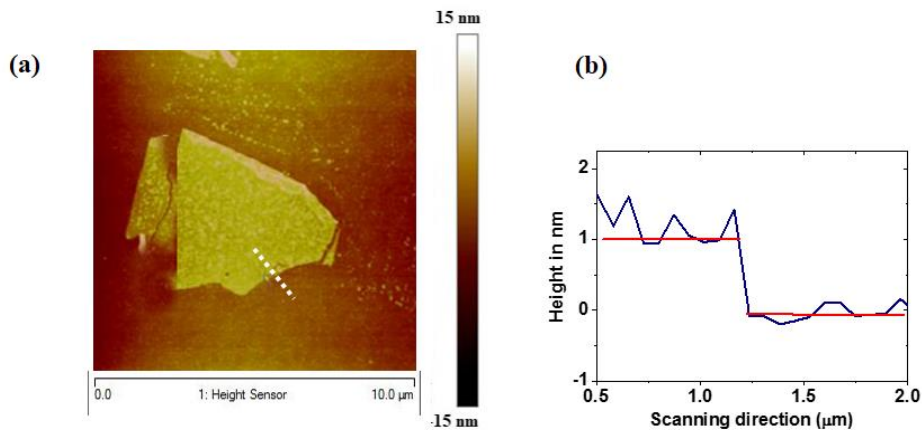

**Figure S1: (a) AFM image and (b) height profile of pristine SL MoS$_2$ flake. The height profile along the cross-section are shown by white dotted line in Figure a.**

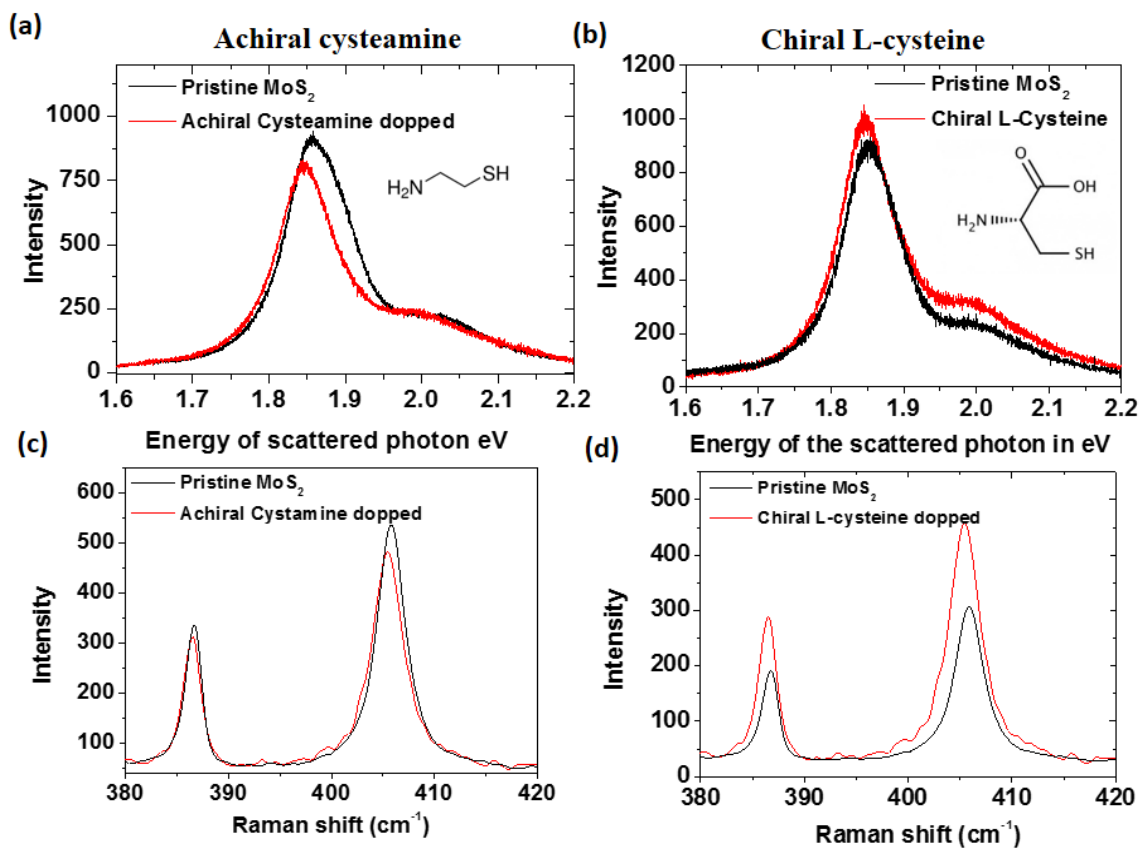

**Figure S2**. Spectral characteristics of MoS$_2$ upon adsorption of achiral/chiral molecules (a,b) compares the photoluminescence spectra, (c,d) represents respective Raman spectra when excited with 2.32eV laser line, and describes change the representative Raman spectra of MoS$_2$ when attached to achiral and chiral molecule. The structure of the corresponding molecules are shown in the inset of Figure S2 (a,b).

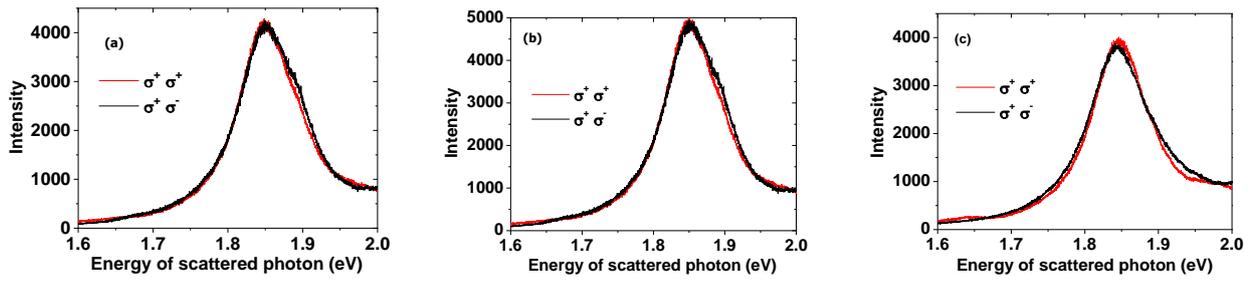

**Figure S3**. Helicity-resolved PL for pristine a) SL-MoS$_2$, b) Ch-SL-MoS$_2$, c) aCh-SL-MoS$_2$ when excited by 2.32eV laser line of left circular polarisation (σ+) at 300K. The detection configuration supports same (σ+) in red and cross (σ-) in black polarisation configuration.